# بررسی تجربی انتقال حرارت نانوسیال در یک لوله مربعی شکل با بار حرارتی ثابت دیواره

**نعسان، توفیق** ؛ زینالی هریس، سعید ؛ نوعی, سید حسین

دانشگاه فردوسی مشهد, دانشکده مهندسی, گروه مهندسی شیمی

## چکیده

در این مطالعه انتقال حرارت جابجایی اجباری نانو سیال آب/ $Al_2O_3$ وآب/ $CuO$ در جریان آرام داخل یک لوله با مقطع مربعی با شار حرارتی ثابت دیواره به صورت تجربی بررسی شده است و تغییرات عدد ناسلت $Nu$ و ضریب انتقال حرارت متوسط بر حسب عدد $Pe$ در غلظتهای مختلف ذرات نانو مورد بررسی قرار گرفته است.

نانوسیال آب/ $Al_2O_3$ با غلظتهای حجمی ذرات نانو ۰،۲%, ۰،۵%, ۱،۰%, ۱،۵%, ۲،۰%, ۲،۵% بررسی شده است. حد اکثر نسبت افزایش ضریب انتقال حرارت جابجایی به ازای هر کدام از غلظتها بررسی شده به ترتیب ۷%, ۱۰%, ۱۳%, ۱۸%, ۲۲%, ۲۷% به دست آمده است. اما نانوسیال آب/ $CuO$ با غلظتهای حجمی ۰،۱, ۰،۲%, ۰،۵%, ۰،۸%, ۱،۰%, ۱،۵% آزمایش شده است. در مقابل حد اکثر افزایش ضریب انتقال حرارت جابجایی در غلظتهای ذکر شده به ترتیب ۸%, ۱۰%, ۱۴%, ۱۶%, ۱۹%, ۲۱% به دست آمده است.

# Experimental investigation of nanofluid heat transfer in square-cross section duct under constant heat flux

Nassan, Taofik; Zeinali Heris, Saeed; Noie, Said Hossin

Ferdowsi University of Mashhad, Faculty of Engineering, Chemical Engineering Department

## Abstract

Forced convective heat transfer of two different nanofluids including $AL_2O_3$/water and CuO/water in laminar flow through square cross section duct under constant heat flux has been investigated. The Nusselt number and average convective heat transfer coefficient for different nanoparticles concentrations as a function of Peclet number have been analyzed.

$AL_2O_3$/water nanofluid with 0.2%, 0.5%, 1.0%, 1.5%, 2.0% and 2.5% volume fractions has been tested. The maximum enhancement of convective heat transfer coefficient for each of the above mentioned concentrations was 7%, 10%, 13%, 18%, 22%, 27%, respectively. Also, CuO/water nanofluid was tested at 0.1%, 0.2%, 0.5%, 0.8%, 1.0% and 1.5% volume fractions and the results show that the maximum enhancement of convective heat transfer coefficient for each concentration was 8%, 10%, 14%, 16%, 19%, 21%, respectively.

# 1- مقدمه

افزایش نجومی فلاکس حرارتی تجهیزات از یک سو و کوچک سازی سیستم های حرارت از سوي دیگر، نیاز به انتقال حرات را در زمان کوتاه و با شدت بالا ضروري مي سازد. مطالعات بسیاري جهت افزایش انتقال حرارت در تجهیزات صورت گرفته است که مي توان به افزایش سطوح حرارتي، ترزیق یا مکش سیال و کاربرد جریان الکتریکي یا میدان مغناطیسي اشاره کرد [1]. روشهاي یاد شده نمي توانند جوابگوي تقاضاي روز افزون انتقال حرارت در فرایندهایي با انتقال انرژي بالا نظیر سیستم هاي لیزري و الکتروني باشند. بنابراین براي افزایش شدت انتقال حرارت به روشهاي جدید نیاز است.

براي بهبود ضریب انتقال حرارت سیالات مطالعات تجربي و تئوریک روي ضریب هدایت حرارتي سیالات حاوي ذرات معلق انجام گرفته است [2]. چوي[3] اولین کسي بود که عبارت نانوسیال را براي سیالات حاوي ذرات معلق نانو به کار برد وبه ضریب هدایت حرارتي بالاي این سیالات اشاره کرد. ایستمن [4] طي مطالعاتي نشان داده که افزایش تقریبي 60درصدي در ضریب هدایت حرارتي آب حاوي 5% حجمي ذرات نانو CuO باشد قابل مشاهده است. میرمعصومي و بهزادمهر [5] با استفاده از مدل مخلوط دوفازي جریان نانوسیال آب/ اکسید آلومینیوم را در یک لوله افقی به صورت عددي مطالعه نموده و تاثیر غلظت نانوذرات روي خواص هیدرودینامیکي و حرارتي سیال را بررسي کردند. نتایج تجربي ژوان و لي [6] نشان دهنده افزایش فوق العاده عدد ناسلت نانوسیال نسبت به آب مي باشد. بررسي هاي کي دي داس و پوترا [7] نشان دهنده تاثیر فوق العاده نانوذرات در فرایند جوشش مي باشد به طوري که با افزایش ذرات نانو جوشش به تعویق افتاده و دماي سطح جوشش افزایش می یابد. زینالي هریس و همکاران [8-10] به طورآزمایشگاهي و عددي انتقال حرارت جابجایي آرام نانوسیالات را مورد بررسي قرار دادند.

با توجه به افت فشار پائیني که کانالهاي مربعي شکل نسبت به لوله هاي مدور ایجاد مي کنند، هدف از این مقاله بررسي انتقال حرارت جابجایي اجباري نانوسیالهاي آب /AL2O3 و آب /CuO و در غلظت ها مختلف نانوذرات در یک کانال مربعي شکل تحت فلاکس حرارتی ثابت دیواره مي باشد تا بتوان جهت کاربردهاي عملي بهترین غلظت ذرات را مورد استفاده قرار گیرد.

# 2- تهیه نانو سیال

در این مطالعه دوسری نانو سیال شامل سوسپانسیون های آب/ $Al_2O_3$ با غلظتهای 0,2%، 0,5%، 1%، 1,5%، 2%، 2,5% و آب/$CuO$ با غلظتهای 0,1%، 0,2%، 0,5%، 0,8%، 1% و 1,5% حجمی ذرات در آب تهیه شده و به منظور بررسی تأثیر نوع و غلظت ذرات در افزایش انتقال حرارت استفاده شده اند.

برای تهیه نانو سیال بعد از تهیه حجم مناسب پودر جامد با توجه به جرم آنها، ذرات نانو با آب مقطر در فلاکس مخلوط می شوند سپس به مدت 8 الی 12 ساعت در همزن اولتراسونیک مدل(Parasonic 3600S ) قرار داده می شوند. بعد از این زمان هیچ ته نشینی برای سوسپانسیون ها در کسر های حجمی مختلف مشاهده نشده است ولی ته نشینی جزئی برای سوسپانسیون 2,5% حجمی ذرات $Al_2O_3$ و سوسپانسیون 1,5% حجمی ذرات $CuO$ در طول مدت آزمایش (4,5 ساعت) مشاهده شد.

# 3- نحوه محاسبات

داده های آزمایشی شامل دمای ورودی و خروجی سیال, دمای دیواره لوله, دبی جریان ونیز مشخصات فیزیکی سیستم به منظور تعیین عدد ناسلت و ضریب انتقال حرارت متوسط نانو سیال در داخل لوله با سطح مقطع مربعی به شکل زیر مورد استفاده قرار می گیرند:

$$\overline{Nu}_{nf}(\exp) = \frac{\bar{h}_{nf}(\exp).D_h}{k_{nf}} \quad (1)$$

$$\bar{h}_{nf}(\exp) = \frac{q}{A.(T_w-T_b)_M} \quad (2)$$

در روابط بالا $D_h$ قطر هیدرولیکی لوله مربعی, $k_{nf}$ ضریب هدایت حرارتی نانو سیال, $q$ فلاکس حرارتی که توسط مقاومت حرارتی تولید شده, $A$ سطح انتقال حرارت , $T_w$ دمای دیواره لوله, $T_b$ دمای بالک, $(T_w-T_b)_M$ اختلاف دمای متوسط,$\overline{Nu}_{nf}(\exp)$ عدد ناسلت متوسط تجربی نانو سیال و $\bar{h}_{nf}(\exp)$ ضریب انتقال حرارت جابجایي متوسط نانو سیال می باشد.

رابطه متداول برای انتقال حرارت در جریان آرام کاملاً توسعه یافته معادله (Seider-Tate) می باشد [11]. نتایج تجربی حاصل از معادله (2) با پیش بینی روابط موجود برای سیال تک فازی داخل لوله با جریان آرام (Seider-Tate) مقایسه می گردد. در این رابطه افزایش انتقال حرارت جابجایي نانو سیال فقط ناشی از افزایش ضریب هدایت گرمایي نانو سیال می باشد. البته باید توجه داشت که خواص مربوط به نانو سیال در داخل معادله استفاده گردد:

ذرات نانو و کسر حجمی ذرات نانو, تأثیر سطح مشترک سیال و ذرات جامد نانو را نیز بر آورد ضریب هدایت حرارتی مد نظر قرار داده است:

$$k_{nf} = \left[\frac{k_s+2k_w+2(k_s-k_w)(1+\beta)^3\emptyset}{k_s+2k_w-(k_s-k_w)(1+\beta)^3\emptyset}\right]k_w \quad (10)$$

در این رابطه $k_w$ ضریب هدایت حرارتی آب, $k_s$ ضریب هدایت حرارتی ذرات نانو و $\beta$ نسبت ضخامت لایه مشترک سیال و ذرات نانو به شعاع ذرات نانو می باشد که در محاسبات انجام گرفته $\beta=0.1$ در نظر گرفته شده است[14].

البته باید توجه داشت که خواص فیزیکی و رئولوژیکی نانوسیال در دمای متوسط بالک تعیین شود.

## 4- سیستم آزمایشگاهی

در این تحقیق برای بررسی انتقال حرارت جابجایی تحت شرایط مرزی با حرارتی ثابت دیواره یک سیستم آزمایشی مطابق شکل (1) ساخته شده است. سیستم آزمایشگاهی شامل یک مدار برای جریان نانو سیال می باشد که دارای بخش های گرمایش با مقاومت حرارتی, سیستم برقی, سرمایش با آب سرد, سیستمهای اندازه گیری ولت, آمپر و دما است. یک مخزن شیشه ای با حجم یک لیتر ونیم مجهز به یک شیر تخلیه به عنوان مخزن ذخیره سیال بکار می رود. به منظور تنظیم دبی جریان عبوری از یک خط برگشت جریان به مخزن استفاده می گردد و با تنظیم شیر موجود بر روی خط برگشتی, دبی مورد نظر به بخش آزمایش ارسال می گردد. همانطور که از شکل (1) مشخص است, بخش آزمایش شامل یک لوله با سطح مقطع مربعی می باشد , این لوله از جنس مس با طول ضلع 1سانتي متر و ضخامت 0.4میلي متر و طول 120سانتي متر می باشد. مقاومت حرارتی11.5 اهم پیرامون 100سانتي متر از لوله مربعی پیچیده شده است. در داخل لوله مربعی نانو سیال در جریان بوده و مقاومت حرارتی با 92.5 ولت DC و 8 آمپر است که فلاکس حرارتی ثابت دیواره را ایجاد می کند.

به منظور کاهش اتلاف حرارتی اطراف مقاومت حرارتی عایق کاری شده است. دو عدد ترموکوپل ( BT-100 ) در بخش ورودی و خروجی سیال به لوله مربعی نصب شده اند تا دمای بالک را اندازه گیری نمایند.

$$\overline{Nu}_{nf}(\text{th}) = 1.86\left(Re_{nf}.Pr_{nf}.\frac{D_h}{L}\right)^{1/3}\left(\frac{\mu_{nf}}{\mu_{wnf}}\right)^{0.14}$$

$(Re_{nf} Pr_{nf} D/L) > 10$ (3)

در رابطه بالا $\left(\frac{\mu_{nf}}{\mu_{wnf}}\right)^{0.14}$ نشانگر تغییرات خواص نانوسیال و تأثیرات جابجایي آزاد می باشد. همچنین $\mu_{nf}$ ویسکوزیته نانوسیال در دمای عملیاتی, $\mu_{wnf}$ ویسکوزیته نانو سیال در دمای دیواره, $Re_{nf}$ , $Pr_{nf}$ و $Pe_{nf}$ به ترتیب عدد رینولدز, پرانتل و پکلت برای نانوسیال می باشند که به شکل زیر تعریف می گردند:

$$Re_{nf} = \frac{\rho_{nf}.\overline{U}.D_h}{\mu_{nf}} \quad (4)$$

$$Pr_{nf} = \frac{C_{Pnf}.\mu_{nf}}{k_{nf}} \quad (5)$$

$$Pe_{nf} = Re_{nf}.Pr_{nf} \quad (6)$$

که $\overline{U}$ سرعت متوسط جریان است, خواص فیزیکی و حرارتی مورد استفاده برای نانو سیال شامل دانسیته, ویسکوزیته, گرمای مخصوص و ضریب هدایت حرارتی با توجه به خواص آب, ذرات نانو و کسر حجمی مورد نیاز در دمای متوسط بالک با استفاده از روابط زیر تعیین می شوند [92]:

$$\rho_{nf} = \emptyset.\rho_s + (1-\emptyset).\rho_w \quad (7)$$

که در آن $\emptyset$ کسر حجمی ذرات نانو, $\rho_s$ دانسیته ذرات نانو و $\rho_w$ دانسیته آب می باشد.

$$\mu_{nf} = \mu_w.(1+2.5\emptyset) \quad (8)$$

معادله (8) برای سوسپانسیون های رقیق نظیر نانو سیال به منظور تعیین ویسکوزیته قابل استفاده است. در این رابطه $\mu_w$ دانسیته آب بوده و این رابطه که به معادله اینشتین معروف است برای غلظت پایین ($\emptyset < 5\%$) برای ذرات کروی صلب قابل استفاده است [12].

گرمای مخصوص نانوسیال با استفاده از رابطه زیر برای سوسپانسیون های جامد-مایع تعیین می گردد:

$$C_{Pnf} = \frac{\emptyset.(\rho_s.C_{Ps})+(1-\emptyset).(\rho_w.C_{Pw})}{\rho_{nf}} \quad (9)$$

در این رابطه$C_{pw}$ گرمايي مخصوص آب و $C_{ps}$ گرماي مخصوص ذرات نانو است.

در بررسی های انجام گرفته معادله چوی-یو[13] تطابق خوبی را با نتایج تجربی برای ضریب هدایت حرارتی نانوسیال نشان داده است[13]. این معادله علاوه بر ضریب هدایت حرارتی سیال و

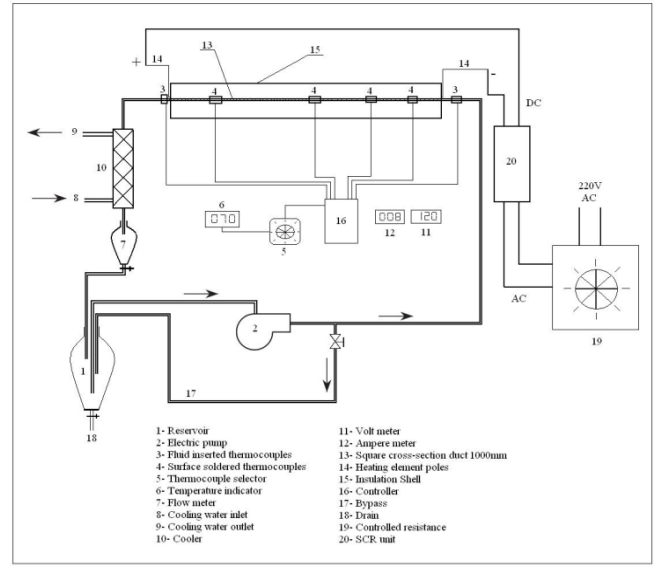

شکل(۱) طرح شماتیک سیستم آزمایشگاهی

چهار عدد ترموکوپل (BT-100) برروی سطح خارجی لوله مربعی با فواصل مختلف به منظور اندازه گیری دمای سطح خارجی لوله مربعی نصب شده اند. سیال توسط پمپ وارد دستگاه اندازه گیری دبی می شود و بعد از آن وارد بخش آزمایش می شود, سپس در یک مبدل حرارتی توسط جریان آب سرد, خنک شده و در نهایت وارد مخزن سیال می شود.

بعد از تهیه نانو سیال با غلظت مشخص و تزریق آن به داخل مخزن شیشه ای, پمپ و جریان آب خنک کن راه اندازی می شوند, سپس مقاومت حرارتی روشن می شود تا فلاکس حرارتی ثابت تولید شود. بعد از ۳۰-۴۰ دقیقه سیستم به حالت پایدار می رسد. شدت جریان سیال با استفاده از شیری که روی خط جریان برگشتی است تنظیم می گردد. آزمایش برای هر غلظت با ۷ دبی مختلف انجام می گیرد. زمانی که دمای سیال خروجی ثابت می ماند اندازه گیری ها شروع می شود.

در طی آزمایش دمای ورودی و خروجی نانو سیال, دمای بدنه لوله در نقاط مختلف و شدت جریان نانو سیال اندازه گیری می شود. هر آزمایش حد اقل دو بار تکرار می گردد.

## ۵- نتایج و بحث:

### ۵-۱ آزمایش با نانوسیال آب / $Al_2O_3$

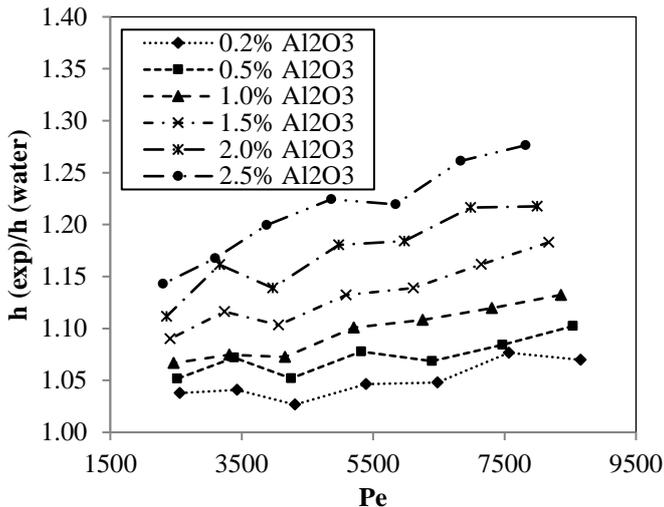

شکل (۲) نسبت ضرایب انتقال حرارت جابجایی نانو سیال آب/$Al_2O_3$ به ضرایب انتقال حرارت جابجایی آب مقطر در غلظتهای مختلف بر حسب عدد پکلت

جهت تعیین میزان افزایش انتقال حرارت توسط نانو سیال در مقایسه با آب خالص نسبت ضریب انتقال حرارت جابجایی نانوسیال $Al_2O_3$/آب به آب مقطر محاسبه شده است. شکل (۲) نشان دهنده افزایش ضریب انتقال حرارت جابجایی نانوسیال $Al_2O_3$/آب در غلظتهای مختلف نسبت به آب می باشد.

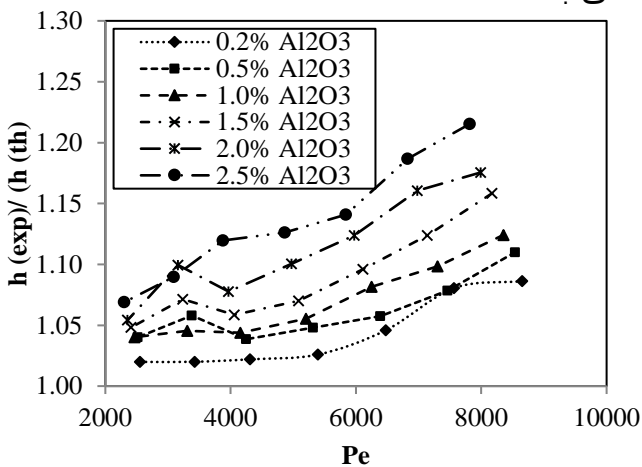

شکل (۳) نسبت ضریب انتقال حرارت تجربی به مقادیر حاصل از معادله Seider-Tate برای نانوسیال $Al_2O_3$/آب در غلظتهای مختلف بر حسب عدد پکلت

به منظور تعیین مقدار و نحوه انحراف معادلات همگن در پیش بینی انتقال حرارت جابجایی نانوسیال, شکل (۳) نسبت ضرایب انتقال حرارت جابجایی نانوسیال $Al_2O_3$/آب را در حالت تجربی به نتایج حاصل از معادلات تئوریک (معادله Seider-Tate) نشان می دهد. این نسبت با افزایش عدد پکلت و نیز کسر حجمی ذرات $Al_2O_3$ افزایش می یابد.

### ۵-۲ آزمایش با نانوسیال آب / CuO

شکل(۴) بیانگر بهبود میزان انتقال حرارت در صورت کاربرد نانو سیال میباشد واین افزایش میزان انتقال حرارت در اعداد پکلت بالا بیشتر است. همچنین با توجه به شکل مشخص می گردد برای کاربردهای عملی نانو سیال آب/CuO با توجه به محدوده این مطالعه غلظت بهینه ۱.۵٪ حجمی خواهد بود.

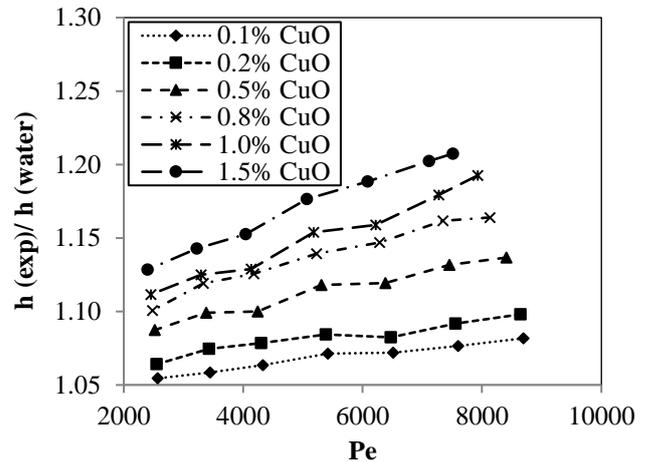

شکل (۴) نسبت ضرایب انتقال حرارت جابجایي نانو سیال آب/CuO به ضرایب انتقال حرارت جابجایي آب مقطر در غلظتهای مختلف بر حسب عدد پکلت

نسبت ضرایب انتقال حرارت تجربی نانو سیال آب/CuO به مقادیر تئوریک حاصل از معادله Seider-Tate در شکل (۵) نشان داده شده است. از شکل مشخص است که این نسبت با افزایش عدد پکلت و نیز غلظت ذرات نانو CuO افزایش می یابد.

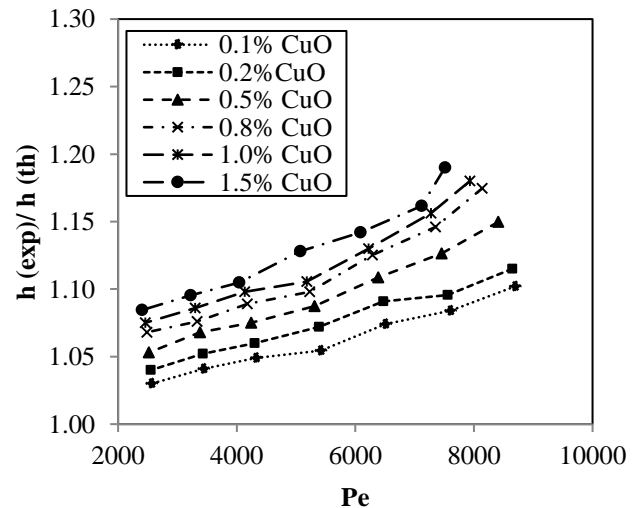

شکل (۵) نسبت ضریب انتقال حرارت تجربی به مقادیر حاصل از معادله Seider-Tate برای نانوسیال آب/CuO در غلظتهای مختلف بر حسب عدد پکلت

## ۵-۳ مقایسه میـزان افـزایش انتقـال حـرارت جابجایي توسط نانوسیال های مختلف

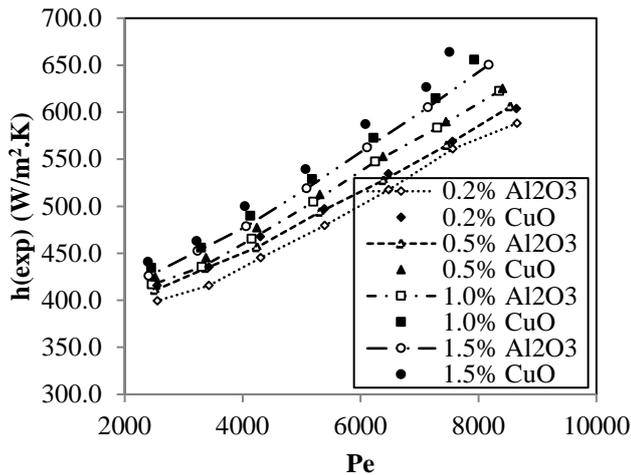

شکل (۶) مقایسه بین ضرایب انتقال حرارت جابجایي نانوسیال های آب/Al2O3 و آب/CuO در غلظتهای مختلف بر حسب عدد پکلت

به منظور مقایسه توانایي انتقال حرارت جابجایي نانوسیال های حاوی ذرات نانو اکسید فلزی, ضرایب انتقال حرارت جابجایي نانوسیال های آب/$Al_2O_3$ و آب/CuO بر حسب عدد پکلت در غلظتهای مختلف (۰٪،۲ ,۰٪،۵ ,۱٪،۰ ,۱٪،۵) در شکل (۶) نشان داده شده اند.

همچنین در شکل (۷) مقایسه اعداد ناسلت نانوسیال های آب/$Al_2O_3$ و آب/CuO در غلظتهای بالا بر حسب عدد پکلت نشان داده شده. شکل نشان می دهد که برای اعداد پکلت کمتر از ۳۷۰۰ اعداد ناسلت نانو سیال اکسید مس در غلظتهای ۱.۰٪ و ۱.۵٪ حتی بالاتر از نانوسیال اکسیدآلومینیوم در غلظت ۲.۵٪ است.

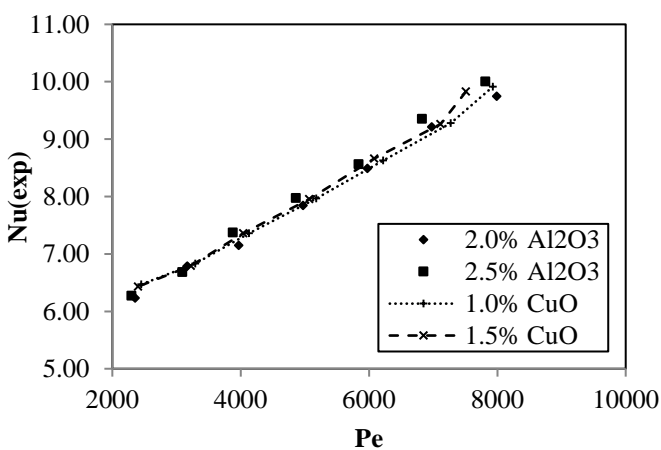

شکل (۷) مقایسه بین اعداد ناسلت نانوسیال های آب/Al2O3 و آب/CuO در غلظتهای بالا بر حسب عدد پکلت

## ۶- نتیجه‌گیری

در این بخش نتایج حاصل از بررسی ها ارائه می گردد:

۱- ضرایب انتقال حرارت جابجایی نانوسیال های اکسید فلزی مورد بررسی با افزایش غلظت ذرات نانو و عدد پکلت در محدوده مورد مطالعه افزایش می یابد.

۲- مقدار افزایش برای تمامی نانو سیال های مورد مطالعه در مقایسه با پیش بینی روابط تئوریک موجود برای انتقال حرارت جریان تک فازی که در آن خواص نانوسیال بکار رفته است بسیار بیشتر است و نشان می دهد که افزایش ضریب انتقال حرارت جابجایی نانوسیال فقط ناشی از افزایش ضریب هدایت حرارتی نبوده بلکه عواملی دیگری در این مورد تأثیر گذار هستند که باید مد نظر قرار گیرند.

۳- نانوسیال حاوی ذرات فلزی CuO بیشترین مقدار افزایش ضریب انتقال حرارت جابجایی در مقایسه با نانوسیال حاوی ذرات فلزی $Al_2O_3$ نشان می دهد.

۴- بطور عمومی افزایش ضریب انتقال حرارت با افزایش هدایت حرارتی نانوسیال که از رابطه یو-چوی محاسبه شده قابل پیش بینی نبود و دلیل این مسئله آن است که عوامل دیگری در افزایش انتقال حرارت نانوسیال تأثیر گذار هستند, که در رابطه اشاره شده در نظر نگرفته است.

۵- افزایش ضریب انتقال حرارت بوسیله نانوسیال جبران کاهش عدد ناسلت و انتقال حرارت بوسیله لوله مربعی را می کند.

# مرجع‌ها